\documentclass[4pt,aps,pre,reprint,superscriptaddress
]{revtex4-2} %
\usepackage[dvipsnames]{xcolor}
\usepackage[utf8]{inputenc}
\usepackage[T1]{fontenc}
\usepackage{amsmath,amssymb,amsthm,bm}

\usepackage{hyperref}

\usepackage{dcolumn}
\usepackage{mathptmx}
\usepackage{amsmath,amssymb,mathtools}
\usepackage{physics}
\usepackage{enumitem}
\usepackage[cal=dutchcal,scr=rsfs,frak=esstix,bb=ams]{mathalfa}

\begin{document}
\title{A microcanonical approach to criticality in the mean-field $\phi^4$ model: evidence of intrinsic microcanonical structure before the thermodynamic limit}
\author{Loris Di Cairano}
\affiliation{Department of Physics and Materials Science, University of Luxembourg, L-1511 Luxembourg City, Luxembourg}
\email{l.di.cairano.92@gmail.com}
\author{Roberto Franzosi}
\affiliation{Department of Physical Science, Earth and Environment, University of Siena, Via Roma 56, 53100 Siena, Italy}
\affiliation{INFN Sezione di Perugia, 06123 Perugia, Italy}
\email{roberto.franzosi@unisi.it}


\date{\today}

\begin{abstract}
Collective critical behavior is often identified with thermodynamic nonanalyticities and divergences emerging only in the infinite-size limit. Here we adopt a complementary viewpoint: criticality is a structural property due to the rearrangement of the interactions among  system's constituents that already exists at finite size and becomes singular only asymptotically. We show that the microcanonical entropy derivatives provide a natural finite-$N$ arena where such structure is encoded in intrinsic extremal/inflection morphologies, and that microcanonical inflection-point analysis (MIPA) turns these morphologies into a unique finite-size critical marker and a well-defined critical trajectory. Using the mean-field $\phi^4$ model as a stringent benchmark, we reconstruct $\beta_N(\varepsilon)$ and $\gamma_N(\varepsilon)$ from microcanonical simulations, validate them against analytic results, and demonstrate that the MIPA trajectory converges to the exact thermodynamic critical point while simultaneously organizing the approach of other observables to their asymptotic behavior. Our results elevate finite-size criticality from a rounded remnant of the thermodynamic limit to a measurable and predictive object in its own right, with direct relevance to modern finite-system platforms and numerical studies.
\end{abstract}

\maketitle

\section{Introduction}

Critical phenomena are traditionally characterized through singular behavior emerging in the thermodynamic limit: nonanalytic thermodynamic observables and scale invariance appearing as $N\to\infty$. While this perspective provides a powerful theoretical framework, it also places the defining features of phase transitions strictly at the level of asymptotic limits. From a physical viewpoint, however, the essential phenomenon behind a phase transition is more elementary and generally size-independent; namely, a reorganization of the interactions among the system's constituents that gives rise to cooperative behaviors that reshape the phase of the system. As external constraints such as energy are varied, the statistical weight in phase space redistributes among competing macroscopic configurations. The thermodynamic singularity is therefore not the phenomenon itself, but rather the limiting expression of an underlying structural reorganization.

This observation suggests a different formulation of criticality. Instead of defining a phase transition through divergences that appear only asymptotically, one may ask whether the collective reorganization itself can be identified directly as a finite-$N$ feature. In the microcanonical ensemble, such a formulation becomes natural because the entropy $s_N(\varepsilon)$ is well defined for any finite system size. Its derivatives with respect to the energy,
\begin{equation}
\beta_N(\varepsilon)=\partial_\varepsilon s_N(\varepsilon), 
\qquad 
\gamma_N(\varepsilon)=\partial_\varepsilon^2 s_N(\varepsilon),
\end{equation}
constitute intrinsic thermodynamic functions at fixed energy. The structure of these derivatives encodes how the distribution of microscopic states reorganizes as the energy constraint is varied. From this viewpoint, criticality can be understood as a specific morphological pattern in the entropy derivatives that exists already at finite $N$ and evolves systematically with system size.

Within this perspective, the notion of a critical point is replaced by that of a \emph{finite-$N$ critical structure}: a robust extremal or inflection pattern appearing in $\beta_N(\varepsilon)$ and $\gamma_N(\varepsilon)$ that signals a collective rearrangement of the phase-space measure. Identifying such structures in a controlled way requires criteria that do not rely on external scaling assumptions or fitting forms. Microcanonical inflection-point analysis (MIPA)~\cite{qi2018classification} addresses this problem by combining a classification of entropy-derivative morphologies with a minimal-sensitivity criterion~\cite{stevenson1981optimized,stevenson1981resolution} that selects stable features under changes in resolution and sampling. This procedure yields a unique finite-$N$ marker and an associated critical trajectory $\varepsilon_\star(N)$ defined entirely from the entropy derivatives themselves.

As we will discuss in Sec.~\ref{sec:state-of-art}, recent studies have explored this approach in several systems where microcanonical entropy derivatives can be computed analytically. These investigations indicate that the derivative structure of the entropy provides a coherent description of the finite-size organization of criticality and its approach to the thermodynamic limit. However, a strong proof of this perspective requires models for which the microcanonical thermodynamics are known analytically, allowing a direct and quantitative comparison between theoretical predictions and simulation-based calculations of the entropy derivatives.

In this work, we provide strong evidence of the strength of the microcanonical approach using the mean-field $\phi^4$ model. This system represents a paradigmatic example of a continuous phase transition with a well-established analytic solution in the thermodynamic limit. We perform microcanonical simulations across a wide range of system sizes and compute the entropy derivatives $\beta_N(\varepsilon)$ and $\gamma_N(\varepsilon)$. The numerical results show quantitative agreement with analytic predictions and reveal a clear morphological signal in the entropy curvature associated with the second-order transition. Applying MIPA to this signal yields a unique finite-$N$ marker and the corresponding critical trajectory $\varepsilon_\star(N)$.

Tracking this trajectory with increasing system size demonstrates a controlled convergence toward the exact critical energy while simultaneously organizing the behavior of independent observables such as magnetization and specific heat. These results support a broader conclusion: the critical phenomenon is not solely encoded in asymptotic singularities but is already present as a structured reorganization of the entropy landscape at finite $N$. In this sense, the thermodynamic-limit singularity emerges as the limiting expression of a sequence of well-defined finite-size critical structures that, although smooth, cannot be relegated to the status of crossover and therefore physically irrelevant.

\section{State of the art: microcanonical criticality and finite-size identification}
\label{sec:state-of-art}

\subsection{From thermodynamic-limit singularities to intrinsic microcanonical structure}

The modern theory of phase transitions was largely formulated in terms of
thermodynamic-limit singularities, through frameworks such as the Yang--Lee
theory of zeros of the partition function and the standard analysis of
nonanalytic free energies and diverging response functions
\cite{yang1952statistical,lee1952statistical}. Within this language,
criticality is primarily identified through structures that become sharp only
as $N\to\infty$.

In parallel, the microcanonical ensemble offers a different viewpoint in which
the central object is the entropy as a function of conserved quantities,
most notably $S(E)=\ln\Omega(E)$. In this setting, the relevant thermodynamic
information is encoded directly in the derivatives of the entropy, such as the
caloric curve and its curvature, rather than in canonical nonanalyticities.
This viewpoint is especially natural for finite and isolated systems, where the
entropy remains well defined independently of the thermodynamic limit.

A major step in this direction was the work of Gross and collaborators, who
emphasized that microcanonical thermodynamics can display sharp and physically
meaningful transition signals already in finite systems
\cite{gross2001microcanonical,gross2000phase,gross2002geometric,gross2005microcanonical}.
In particular, they showed that phase coexistence and phase separation may
appear through convex regions of the entropy and negative microcanonical heat
capacity, features that are suppressed or re-encoded in canonical descriptions
\cite{gross2005microcanonical,gross2001microcanonical}. These works helped
establish that microcanonical signals are not merely finite-size artifacts,
but can reflect genuine structural properties of many-body organization.

\subsection{Ensemble inequivalence, long-range interactions, and the role of mean-field models}

The conceptual status of microcanonical anomalies became sharper with the increasing observations of ensemble inequivalence, where nonconcave entropy implies nonequivalence at the thermodynamic level and metastability at the macrostate level
\cite{touchette2003equivalence,touchette2004introduction,touchette2005nonequivalent,ellis2002nonequivalent,ellis2004thermodynamic}.
This framework clarified why features such as negative specific heat,
temperature jumps, and anomalous susceptibilities can be thermodynamically
consistent rather than pathological.

Long-range interacting systems provide a particularly important arena in this
respect, since inequivalence is often generic rather than exceptional. In this regard, many works highlighted the central role of solvable mean-field models in the
study of nonequivalent ensembles, negative specific heat, and unusual relaxation phenomena~\cite{campa2014physics,barre2001inequivalence,leyvraz2002ensemble,Pikovsky2014EnsembleInequivalence}.
Related developments include the treatment of negative-temperature regimes
\cite{miceli2019statistical} and microcanonical formulations in gravitational
and high-energy contexts \cite{brown1993microcanonical,strominger1983microcanonical,Hagedorn1965}.

For the present work, this body of literature is relevant for two reasons.
First, it shows that the microcanonical ensemble is not merely an alternative
representation, but often the natural framework in which the underlying
thermodynamic structure becomes visible. Second, it establishes solvable
mean-field models as a privileged benchmark class for testing new microcanonical criteria.

\subsection{Finite-size systems and structural signals beyond the thermodynamic limit}

Independently of long-range physics, the study of finite systems in nuclear
fragmentation, clusters, and related contexts motivated the search for
transition criteria that remain meaningful away from the thermodynamic limit. In this direction, it was shown that finite-size systems can exhibit robust signatures of transition-like behavior, and that such behavior may be formulated consistently without invoking asymptotic singularities
\cite{chomaz2006challenges,chomaz1999energy,gulminelli1999critical}.

A conceptually important development was the observation that, even for smooth Hamiltonians, derivatives of entropy may display many nonanalyticities already at finite $N$, but the thermodynamic limit usually washes most of them out, leaving just the infinite-size nonanalyticity which is then identified as the real transition~\cite{casetti2006nonanalyticities,CasettiKastnerNerattini2009}. This is the case of the 1D mean-field Berlin-Kac model whose result
sharpened an essential distinction: the existence of structural features in
$S_N(E)$ at finite size is not by itself equivalent to the existence of a
macroscopic phase transition, but neither can such features be dismissed as
irrelevant. More concretely, this result shows that the identification of phase transitions through nonanalyticities cannot be used at finite-size since, in general, this will not yield a phase transition in the thermodynamic limit. In this perspective, we have shown in Ref.~\cite{di2026criticality} that the
breakdown of finite-$N$ nonanalyticity as a transition criterion does not
invalidate finite-size criticality itself. Rather, it forces a shift of
focus: from nonanalytic points to stable morphological structures in the
microcanonical entropy derivatives. Within this framework, criticality is
understood as an intrinsic finite-$N$ organization already encoded in
$\beta_N(E)$ and $\gamma_N(E)$, while the thermodynamic singularity appears
only as the asymptotic limit of this underlying sequence of finite-size
structures.

This distinction is directly relevant to the present paper, where
the goal is precisely to identify finite-$N$ critical structures whose
systematic evolution leads to the thermodynamic singularity.

\subsection{Microcanonical inflection-point analysis and finite-size critical markers}

Among the approaches specifically aimed at extracting transition signals from
finite-size microcanonical data, the most directly relevant one is microcanonical inflection-point analysis (MIPA). Introduced by Bachmann et al. \cite{schnabel2011microcanonical}, MIPA shifts the focus from singularities to the morphology of successive entropy derivatives. The central idea is that cooperative behavior in finite systems
can be identified through robust inflection or extremal structures in $S^{(k)}(E)$, rather than through thermodynamic-limit nonanalyticities.

Bachmann and collaborators later generalized this framework by combining it with the
principle of minimal sensitivity, thereby formulating a systematic classification scheme for phase transitions of arbitrary order in terms of stable entropy-derivative morphologies \cite{stevenson1981optimized,stevenson1981resolution,qi2018classification}.
Subsequent applications extended this program to a broader set of systems,
including Ising strips and chains, aggregation phenomena, and higher-order
transition-like structures
\cite{sitarachu2020exact,koci2017subphase,sitarachu2022evidence,rocha2025microcanonical}. MIPA is now applied to a wide range of physical systems, such as the Potts model~\cite{Wang2024PottsFiniteSize,Liu2025PottsGeometry}, $Z(N)$-clock models~\cite{Shi2026SixStateClock}, and other spin systems~\cite{Liu2022IsingBaxterWu,Liu2025BlumeCapel}, as well as in lattice field theories~\cite{bel2021geometrical,di2022geometrictheory}, glass-transitions~\cite{vesperini2025glass}, and long-range $(1/r^\alpha)$ interacting systems~\cite{di2025geometric,di2025phase}.

This line of work is the closest antecedent of the present study. It provides the operational language in which one may speak of finite-$N$ critical structure without reducing it either to a finite-size pseudocritical fit or to a mere precursor of the thermodynamic limit.

\subsection{Positioning of the present work}

Within this landscape, the present work is positioned at the intersection of
three ideas. First, the microcanonical ensemble provides the natural setting
in which finite-size critical structure can be defined through entropy
derivatives. Second, solvable mean-field systems offer stringent benchmarks
because their thermodynamic-limit behavior is analytically accessible.
Third, MIPA supplies the operational criterion that turns entropy-derivative
morphology into a unique finite-$N$ marker and hence into a critical
trajectory.

The specific contribution of this paper is to implement this program in the
mean-field $\phi^4$ model, a benchmark where the thermodynamic-limit solution
is known analytically. By combining finite-$N$ microcanonical reconstruction of $\beta_N(\varepsilon)$ and $\gamma_N(\varepsilon)$ with the exact infinite-size reference curves, we show that the finite-size extremal structure selected by MIPA converges systematically toward the exact critical energy. In this way, such work aims to strengthen the claim that criticality can be formulated as an intrinsic finite-size property through the morphology of microcanonical entropy derivatives.

\section{Mean-field $\phi^4$ model as a benchmark for intrinsic finite-size criticality}

\subsection{Model and thermodynamic-limit behavior}

We consider the mean-field $\phi^4$ model defined by the Hamiltonian~\cite{campa2007negative}:
\begin{equation}
H = \sum_{i=1}^{N} \left(\frac{p_i^2}{2} - \frac{1}{4}\phi_i^2 + \frac{1}{4}\phi_i^4 \right)
- \frac{J}{4N}\left(\sum_{i=1}^{N}\phi_i\right)^2 .
\end{equation}
The variables $\phi_i$ represent scalar degrees of freedom coupled through a 
mean-field ferromagnetic interaction of strength $J$. 
This system provides a paradigmatic example of a continuous symmetry-breaking transition in a mean-field setting.

In the thermodynamic limit, the model can be exactly solved \cite{campa2007negative} and exhibits a second-order phase transition
separating a high-energy disordered phase from a low-energy phase with spontaneous
magnetization. The critical point can be determined analytically within the canonical solution of the model, yielding a critical temperature $T_c \simeq 0.2645$ and a corresponding critical energy density $\varepsilon_c \simeq 0.132$.
Because the thermodynamic properties of the model are known with high precision,
the system provides an ideal reference for testing finite-size reconstructions
of thermodynamic quantities.

Previous studies have used this model primarily to investigate the geometry
of the entropy surface $s(\varepsilon,m)$ \cite{campa2006microcanonical,hahn2005mean} and the resulting inequivalence between ensembles~\cite{campa2007negative}, including regimes of negative magnetic susceptibility.
Here, we adopt a complementary viewpoint: rather than focusing on the entropy as a function of multiple macroscopic variables, we examine the structure of the entropy derivatives with respect to the energy alone. In particular, our goal is to test whether critical behavior is already encoded at finite system size and can be identified directly from the morphology of the microcanonical entropy derivatives. 

As in other mean-field models, such as the Berlin-Kac and XY models, the thermodynamic-limit solution of the mean-field $\phi^4$ model is known analytically, providing a precise reference for the location and nature of the transition. Second, the continuous nature of the transition makes the system an ideal test bed for identifying subtle finite-size structures in the entropy curvature that precede the thermodynamic singularity.

\section{Statistical mechanics in Canonical and Microcanonical ensembles}

In this work, the model is analyzed through three complementary levels of description.
At finite system size, we perform numerical simulations in both the microcanonical and
canonical ensembles. The microcanonical simulations provide direct access to the entropy
derivatives, while the canonical simulations yield the standard thermodynamic observables,
such as the caloric curve, the specific heat, and the magnetization, as functions of the
temperature. In addition, the thermodynamic-limit behavior is obtained independently from
the canonical mean-field solution, which provides the exact infinite-size reference curves
used throughout the paper.

In the microcanonical ensemble, the fundamental thermodynamic quantity is the specific entropy
\begin{equation}
s_N(\varepsilon)=\frac{1}{N}\ln \Omega_N(\varepsilon),
\end{equation}
where $\Omega_N(\varepsilon)$ denotes the density of states at energy density
$\varepsilon=E/N$, defined by
\begin{equation}
\Omega_N(\varepsilon)
=
\int \delta\!\left(H(p,q)-E\right)\,d^Nq\,d^Np .
\end{equation}
Its derivatives are intrinsic finite-size observables. The first derivative
\begin{equation}
\beta_N(\varepsilon)=\partial_\varepsilon s_N(\varepsilon),
\end{equation}
is the microcanonical inverse temperature, while the second derivative
\begin{equation}
\gamma_N(\varepsilon)=\partial_\varepsilon^2 s_N(\varepsilon),
\end{equation}
measures the local curvature of the entropy and is related to the microcanonical
specific heat through
\begin{equation}
C_v(\varepsilon)=-\frac{\beta_N(\varepsilon)^2}{\gamma_N(\varepsilon)}.
\end{equation}
From the viewpoint adopted here, these derivatives are the primary quantities of
interest: rather than identifying the transition through thermodynamic-limit singularities,
we examine the morphology of $\beta_N(\varepsilon)$ and $\gamma_N(\varepsilon)$ themselves,
whose extrema and inflection structures encode the collective reorganization of the system.

To estimate these quantities at finite $N$, we perform microcanonical simulations and
compute the entropy derivatives using the Pearson--Halicioglu--Tiller
formalism~\cite{pearson1985laplace}. Writing $k:=K/N$ for the specific kinetic energy,
one has
\begin{equation}
\label{eq:I_II_derMicroCanS_rewritten}
\begin{split}
\partial_{\varepsilon}s_N(\varepsilon)
&=
\left(\frac{1}{2}-\frac{1}{N}\right)\langle k^{-1}\rangle_{\varepsilon},
\\
\partial^{2}_{\varepsilon}s_N(\varepsilon)
&=
N\Bigg[
\left(\frac{1}{2}-\frac{1}{N}\right)
\left(\frac{1}{2}-\frac{2}{N}\right)
\langle k^{-2}\rangle_{\varepsilon}
\\
&\qquad\qquad
-
\left(\frac{1}{2}-\frac{1}{N}\right)^2
\langle k^{-1}\rangle_{\varepsilon}^{\,2}
\Bigg].
\end{split}
\end{equation}
These relations provide nonperturbative finite-size estimators of the entropy
derivatives directly in the microcanonical ensemble. Details of the sampling
procedure and numerical implementation are reported in Refs.~\cite{di2025geometric,di2025phase}.

In parallel, we also perform canonical (Monte Carlo) simulations at finite $N$ based on computing the canonical observables through temperature derivatives of the (specific) free energy:
\begin{equation}
    f_N(T)=-\frac{T}{N}\,\log Z_N(T)
\end{equation}
\begin{equation}
    Z_N(T):=\int e^{-H(p,q)/T}\;d^N q\,d^Np\,,
\end{equation}
on the partition function. Note that, in order to have direct comparison with the microcanonical-based estimation, we use the Hamiltonian and not simply the potential function.

Then, the caloric curve can be written as
\begin{equation}
    \varepsilon_N(T)=f_N(T)-T\frac{\partial f_N}{\partial T}=\frac{\langle H\rangle_T}{N}.
\end{equation}
and, a further differentiation with respect to temperature yields the specific heat:
\begin{equation}
    c_{V,N}(T)=\frac{\partial \varepsilon_N}{\partial T} =
\frac{1}{N T^2}
\left(
\langle H^2\rangle_T-\langle H\rangle_T^2
\right).
\end{equation}

These quantities provide the standard finite-size thermodynamic signatures of the
transition and serve as an independent level of description with which the
microcanonical results can be compared.

The thermodynamic-limit reference curves are obtained independently from the canonical
mean-field solution. In the limit $N\to\infty$, the model reduces to an effective
one-site problem parametrized by the magnetization
\begin{equation}
m=\frac{1}{N}\sum_i \phi_i .
\end{equation}
At fixed temperature $T=\beta^{-1}$, the equilibrium branch $m^\star(T)$ is determined by
the self-consistency equation
\begin{equation}
    m =\frac{\int d\phi\,\phi\,\exp\!\left[-\beta v(\phi)+\frac{\beta m}{4}\phi\right]}{\int d\phi\,
    \exp\!\left[-\beta v(\phi)+\frac{\beta m}{4}\phi\right]},
\end{equation}
with local potential
\begin{equation}
    v(\phi)=-\frac{\phi^2}{4}+\frac{\phi^4}{4}.
\end{equation}
Once $m^\star(T)$ is known, the canonical energy density follows as
\begin{equation}
    \varepsilon(T)= \frac{T}{2} + \langle v(\phi)\rangle_{m^\star,T}-\frac{(m^\star)^2}{4},
\end{equation}
where $\langle\cdot\rangle_{m^\star,T}$ denotes the expectation value with
respect to the effective one-site measure. This yields the exact infinite-size
caloric curve in parametric form, and the canonical specific heat is obtained as
\begin{equation}
C_V(T)=\frac{d\varepsilon(T)}{dT}.
\end{equation}

The critical temperature follows from linearizing the self-consistency equation
around the symmetric branch $m=0$. Expanding for small $m$ gives
\begin{equation}
\langle \phi\rangle_{m,T}
=
\frac{m}{2T}\,\langle \phi^2\rangle_{0,T}
+O(m^3),
\end{equation}
which leads to
\begin{equation}
\chi_0(T_c)=2T_c,
\qquad
\chi_0(T)
=
\frac{\int d\phi\,\phi^2 e^{-v(\phi)/T}}
{\int d\phi\, e^{-v(\phi)/T}}.
\end{equation}
At the critical point, $m^\star(T_c)=0$, so that the corresponding critical energy density is
\begin{equation}
\varepsilon_c
=
\frac{T_c}{2}
+
\langle v(\phi)\rangle_{0,T_c}.
\end{equation}

The canonical thermodynamic-limit solution also provides direct access to the
corresponding microcanonical observables. Along the equilibrium branch, the canonical
and microcanonical temperatures coincide, so that the caloric curve can be inverted to define
\begin{equation}
\beta_\infty(\varepsilon)
=
\frac{\partial s_\infty(\varepsilon)}{\partial \varepsilon}
=
\frac{1}{T_\infty(\varepsilon)}.
\end{equation}
Differentiating once more with respect to the energy yields the entropy curvature
\begin{equation}
\gamma_\infty(\varepsilon)
=
\frac{\partial^2 s_\infty(\varepsilon)}{\partial \varepsilon^2}
=
\frac{d\beta_\infty}{d\varepsilon}.
\end{equation}
In practice, the inversion is conveniently performed separately on the ordered
and disordered branches around $\varepsilon_c$, where the caloric curve changes slope.
This construction yields the exact infinite-size reference curves
$\beta_\infty(\varepsilon)$ and $\gamma_\infty(\varepsilon)$, which are then compared
one-to-one with the finite-$N$ estimates obtained from
Eq.~\eqref{eq:I_II_derMicroCanS_rewritten}.

The analysis that follows is therefore based on a threefold comparison:
finite-$N$ canonical observables, finite-$N$ microcanonical entropy derivatives,
and the exact thermodynamic-limit reference curves.
In the next section, we first discuss the standard canonical quantities,
namely the caloric curve and the specific heat, and then turn to the
microcanonical entropy derivatives, where the intrinsic finite-$N$
critical structure becomes directly visible.

\section{Entropy-derivative morphology and the emergence of a finite-$N$ critical structure}
\label{sec:mipa-results}

\begin{figure*}
    \centering
    \includegraphics[width=0.9\linewidth]{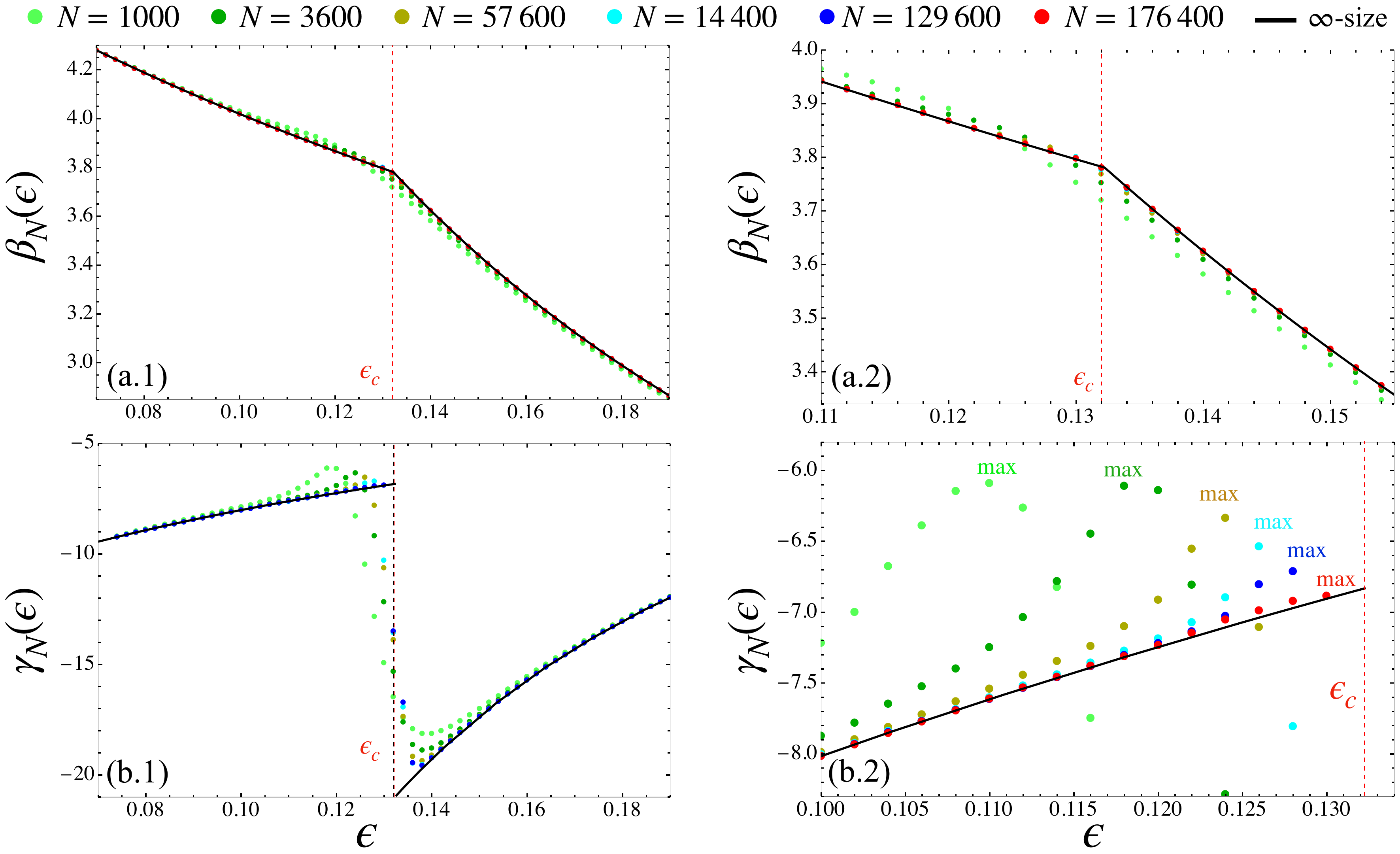}
    \caption{Microcanonical entropy derivatives for the mean-field $\phi^4$ model. (a) Inverse temperature $\beta_N(\varepsilon)$ reconstructed from microcanonical simulations for several system sizes. (b) Entropy curvature $\gamma_N(\varepsilon)=\partial_\varepsilon^2 s_N(\varepsilon)$. While $\beta_N(\varepsilon)$ varies smoothly across the critical region, the curvature $\gamma_N(\varepsilon)$ develops a localized extremal structure that becomes progressively sharper with increasing $N$. This structure reflects a local restructuring of the entropy landscape associated with the onset of critical behavior. The position of this extremal feature defines the finite-size critical marker $\varepsilon_\star(N)$ through microcanonical inflection-point analysis (MIPA).
}
    \label{fig:entropy_derivatives}
\end{figure*}

\subsection{Microcanonical classification of phase transitions}

As discussed above, from canonical-based calculations, we know that the 1D mean-field $\phi^4$ model undergoes a second-order phase transition. In order to properly analyze such a transition in the microcanonical ensemble, we employ the microcanonical inflection point analysis (MIPA) and, in particular, the definition for second-order transition. According with Ref.~\cite{qi2018classification}, a second-order transition occurs at $\epsilon_c=E_c/N$ if and only if $\beta_N:=\partial_E s_N$ admits an inflection point at $\epsilon=\epsilon_c$ such that $\gamma_N:=\partial^2_\epsilon s_N$ has a negative-valued (local) maximum at $\epsilon=\epsilon_c$, i.e., $\gamma_N(\epsilon_c)<0$.

Then, we analyze the microcanonical observables $\beta_N$ and $\gamma_N$ obtained from microcanonical simulations and look for the MIPA-pattern formulated above. Figure~\ref{fig:entropy_derivatives}(a) shows the finite-size $\beta_N(\varepsilon)$ (colored disks) against the exact infinite-size curve $\beta_\infty(\epsilon)$ (continuous black curve) for several system sizes across the energy range containing the critical point. The curves display a smooth and well-resolved structure whose overall behavior is consistent with the analytic thermodynamic-limit prediction.

As the system size increases, the slope of $\beta_N(\varepsilon)$ near the critical region becomes progressively steeper, reflecting the approach toward the thermodynamic singularity.
Outside the critical region, the curves for different $N$ collapse rapidly onto a common envelope, indicating that the intrinsic structure of criticality is the inflection point as it converges to a kink in the thermodynamic limit.

Figure~\ref{fig:entropy_derivatives}(b) shows the second derivative $\gamma_N(\varepsilon)$ for the same set of system sizes. In contrast to $\beta_N$, the behavior of $\gamma_N$ is sharper, making it easier to identify relevant patterns around the critical region. In fact, for each system size, $\gamma_N(\varepsilon)$ (colored disks in panel~\ref{fig:entropy_derivatives}(b.1)) exhibits a positive-valued maximum (pronounced peak) whose position moves systematically with $N$. Focusing on panel~\ref{fig:entropy_derivatives}(b.2), as the system size increases, $\gamma_N$ tends to overlap the thermodynamic-limit curve $\gamma_\infty$ (see the continuous black curve). Interestingly, these maxima tend to get closer and closer to the maximum value of the thermodynamic-limit curve, which coincides with the jump at $\epsilon=\epsilon_c$.

In conclusion, this analysis leads to a crucial result. Criticality is an intrinsic property of the system, independent of the size of the system, and therefore not a property of the thermodynamic limit. The signature of a phase transition is already present in the (microcanonical) thermodynamic observables, and what matters is not the way we extrapolate to infinite-size but the way we read them independently of the size. This consists of identifying specific morphological structures, such as inflection points and maxima, as predicted by MIPA. Apparently, these rules seem to detect crossovers or finite-size effects that have nothing to do with real phase transitions. This analysis shows that such maxima are the seed of the transition, as for increasing system sizes, they converge to the exact asymptotic curve (panel~\ref{fig:entropy_derivatives}(b.2)).

\subsection{Identification of the finite-$N$ critical marker}

The maximum-based signal observed in $\gamma_N(\varepsilon)$ provides the basis for identifying the finite-size critical structure of the system. Applying this criterion to the curvature curves shown in Fig.~\ref{fig:entropy_derivatives}(b) yields a unique energy value for each system size. This value defines the finite-$N$ critical marker $\varepsilon_\star(N)$, which specifies the location of the entropy-derivative structure
associated with the critical reorganization.

Importantly, this marker is defined directly from the morphology of the entropy derivatives themselves and does not rely on external fitting procedures or scaling assumptions. The resulting sequence $\varepsilon_\star(N)$ therefore represents an intrinsic characterization of the critical behavior at finite system size. The existence of a well-defined marker $\varepsilon_\star(N)$ naturally leads to a further question: whether the sequence of finite-size structures identified in this way converges in a systematic manner toward the thermodynamic critical point.

To address this question, we now examine the behavior of the critical trajectory $\varepsilon_\star(N)$ as the system size
is increased.

\section{Critical trajectory and convergence toward the thermodynamic limit}

\begin{figure}
    \centering
    \includegraphics[width=0.9\linewidth]{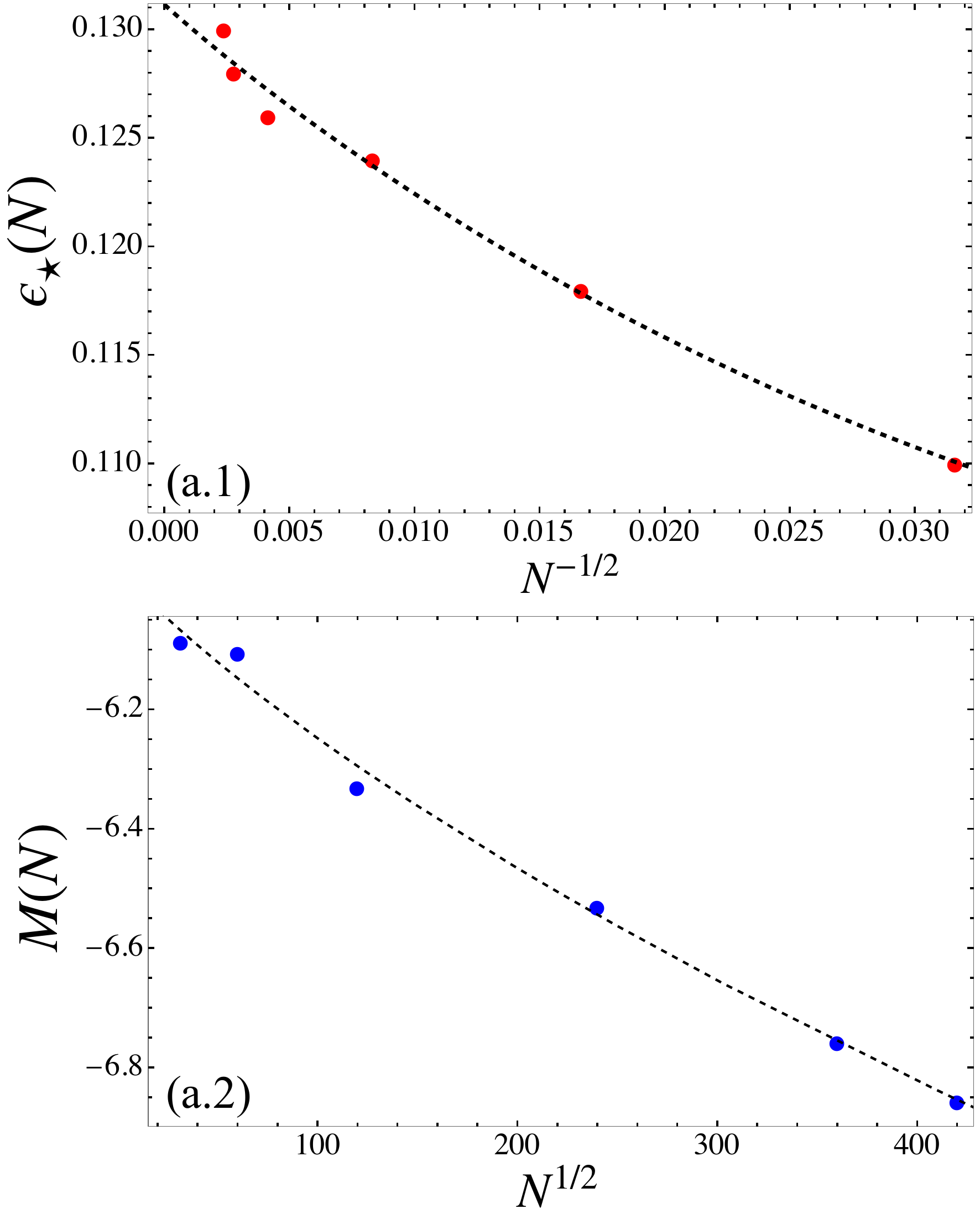}
    \caption{Finite-size critical trajectory extracted from the entropy curvature. (a) Energy values $\varepsilon_\star(N)$ identified from the extremal structure of $\gamma_N(\varepsilon)$ for all simulated system sizes. (b) Convergence of $\varepsilon_\star(N)$ toward the thermodynamic critical energy $\varepsilon_c$ as a function of $N^{-1/2}$. The dashed line indicates the exact critical energy. The data reveal a smooth and systematic finite-size flow toward $\varepsilon_c$, showing that the thermodynamic critical point emerges as the limit of a well-defined family of finite-size entropy structures.}
\label{fig:critical_trajectory}
\end{figure}

\subsection{Definition of the finite-size critical trajectory}

The identification of the finite-$N$ marker $\varepsilon_\star(N)$ naturally defines a sequence of energies that track the location of the entropy-derivative structure as the system size varies.

Figure~\ref{fig:critical_trajectory} panel (a.1) shows the resulting values of $\varepsilon_\star(N)$ obtained from the curvature signal $\gamma_N(\varepsilon)$ for all simulated system sizes, while panel~\ref{fig:critical_trajectory}(a.2) represents the value of the maxima at the critical energy value, namely, $M(N):=\gamma_N(\epsilon_\star(N))$. Rather than appearing as isolated pseudocritical estimates, these values form a trajectory in the $(N^{-1/2},\varepsilon)$ and $(N^{-1/2},\max_\epsilon\gamma_N)$ planes, respectively. This trajectory represents the finite-size evolution of the entropy-derivative structure identified in Sec.~\ref{sec:mipa-results}. Each point corresponds to the energy at which the microcanonical entropy curvature signals the collective reorganization of the phase-space measure for that system size.

\subsection{Approach to the thermodynamic critical point}

To analyze the convergence of the finite-size critical trajectory we exploit the method introduced in Ref.~\cite{di2026criticality}. Figure~\ref{fig:critical_trajectory}(a.1) displays $\varepsilon_\star(N)$ as a function of $N^{-1/2}$ and analogously for $M(N)$ as a function of $N^{-1/2}$.
The data exhibit a systematic approach toward the
known thermodynamic critical energy $\varepsilon_c$.

This behavior indicates that the entropy-derivative structures identified at finite system size evolve continuously into the thermodynamic singularity.
The critical point of the infinite system, therefore, emerges as the limit of a well-defined family of finite-size critical structures.

In this sense, the thermodynamic singularity does not appear abruptly at $N\to\infty$, but rather represents the endpoint of an ordered sequence of reorganizations already present in finite systems.

\subsection{Thermodynamic criticality as the limit of finite-$N$ structures}

The existence of the trajectory $\varepsilon_\star(N)$ provides a direct connection between finite-size behavior and thermodynamic criticality.

Within the conventional framework, the finite system displays only rounded anomalies whose interpretation relies on extrapolation to the thermodynamic limit. In contrast, the entropy-derivative analysis reveals that the finite system possesses a well-defined structural marker of the critical reorganization. As the system size increases, this marker evolves smoothly and
converges toward the thermodynamic critical point.
From this perspective, the singular behavior of the infinite system can be viewed as the limiting manifestation of a sequence of finite-size entropy structures.

This viewpoint emphasizes that criticality is not solely a property of asymptotic limits but corresponds to an organized restructuring of the entropy landscape that is already present at finite $N$. The intrinsic trajectory $\varepsilon_\star(N)$ extracted from the
entropy derivatives therefore provides a natural reference for organizing the behavior of other thermodynamic observables.

In the following section, we analyze the system from the canonical ensemble, along with independent quantities such as the magnetization and Binder-type cumulants, which exhibit their characteristic finite-size evolution around this entropy-based critical trajectory.

\section{Standard finite-size thermodynamic signatures and their limitations}

\subsection{Caloric curve and specific heat}

At this stage, it is instructive to consider how the approach to criticality appears in thermodynamic observables. In the canonical ensemble, finite-size signatures of a second-order phase transition usually manifest through peaks in the specific heat that transform into nonanalyticities in the thermodynamic limit. 

\begin{figure*}
    \centering
    \includegraphics[width=0.9\linewidth]{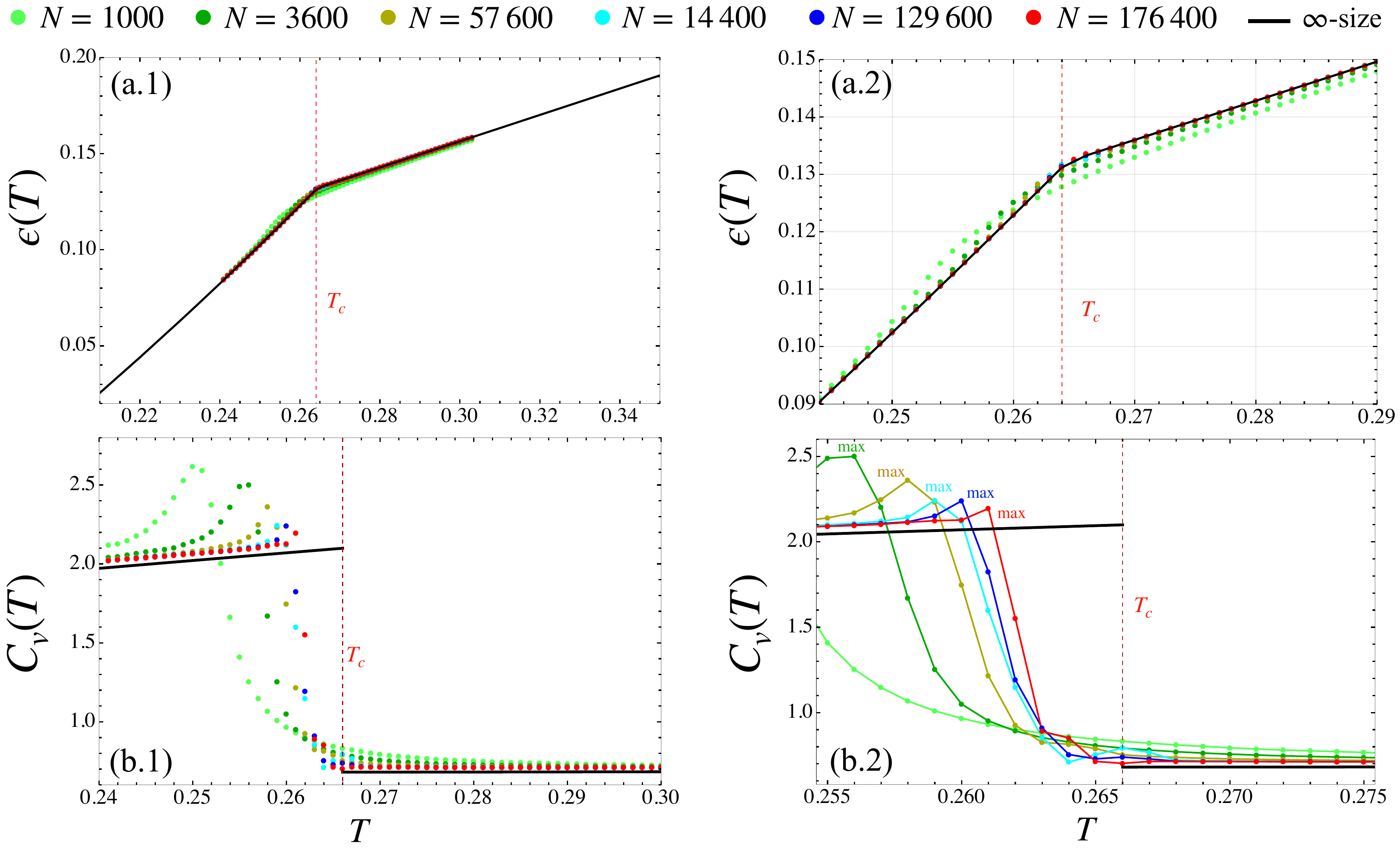}
    \caption{Canonical finite-size signatures of the phase transition in the mean-field $\phi^4$ model.(a) Caloric curve $\varepsilon(T)$ for several system sizes. (b) Specific heat $C_V(T)$ obtained from canonical simulations. The vertical dashed line indicates the exact critical temperature of the thermodynamic limit. As expected for a continuous transition, the specific heat develops a size-dependent peak that sharpens and shifts with increasing $N$. At finite system size the thermodynamic singularity is replaced by smooth anomalies whose position defines a pseudocritical temperature. These standard observables clearly signal the approach to criticality, yet they do so through broadened and size-dependent features. This behavior motivates the search for an intrinsic finite-size description based directly on the entropy landscape.}
    \label{fig:canonical_signatures}
\end{figure*}

Figure~\ref{fig:canonical_signatures} shows the caloric curve $\varepsilon(T)$ (panels~\ref{fig:canonical_signatures}(a.1) and (a.2)) and the corresponding specific heat $C_v(T)$ (panels~\ref{fig:canonical_signatures}(b.1) and (b.2)) for increasing system sizes. Here, we also report the thermodynamic-limit curve for comparison, namely, the continuous black curve.
The vertical dashed red line indicates the exact critical temperature, $T_c$, of the thermodynamic limit. 

As expected for a continuous phase transition, the specific heat
develops a pronounced peak (maximum) near the critical region, see panel~\ref{fig:canonical_signatures}(b.1). From standard real analysis, this is due to the presence of inflection points in $\epsilon_N(T)$ that, after differentiation ($\partial_T\epsilon(T)\equiv C_v(T)$), produce such maxima. Focusing on the caloric curve in panel~\ref{fig:canonical_signatures}(a.1), we see that the infinite-size $\epsilon_\infty(T)$ admits a cusp-like behavior at $T_c$ (see the continuous black line). Zooming on the critical region (see panel~\ref{fig:canonical_signatures}(a.2)), we observe the convergence of the finite-size $\epsilon_N$ to $\epsilon_\infty$. The caloric curve, in fact, shows a progressively sharper change
of slope as the system size increases, reflecting the emergence
of the thermodynamic singularity. This suggests that the inflection points (responsible for the occurrence of the peaks in the specific heat) are the finite-size signature of the transition. 

The presence of the cusp in $\epsilon_\infty$ is the source of nonanalyticities in the specific heat which indeed admits a jump in correspondence with $T_c$; see the continuous black line in panel~\ref{fig:canonical_signatures}(b.1). However, at finite system sizes, the anomaly remains smooth, but the shifting of the peaks' position with increasing $N$ suggests a clear critical trajectory.

\section{Discussion and conclusions}

Criticality is most often introduced through the singular structures that appear only in the thermodynamic limit. Within that viewpoint, finite systems display only rounded anomalies whose meaning is fixed retrospectively, through extrapolation toward $N\to\infty$. The analysis developed here supports a different interpretation. In the mean-field $\phi^4$ model, the finite system already possesses a well-defined structural signal of the transition, and this signal is encoded directly in the morphology of the microcanonical entropy derivatives. The thermodynamic singularity is therefore not the definition of the phenomenon but its asymptotic limit.

At the canonical level, the caloric curve and the specific heat show the familiar finite-size pattern of a continuous transition: smooth inflection structures in $\varepsilon_N(T)$, broadened peaks in $c_{V,N}(T)$, and a systematic drift toward the exact critical temperature. These observables clearly signal the approach to criticality, but they do so indirectly, through size-dependent anomalies whose interpretation remains tied to the thermodynamic limit. Taken by themselves, they do not provide an intrinsic finite-$N$ notion of the transition.

At the microcanonical level, the picture changes qualitatively. The reconstructed $\beta_N(\varepsilon)$ and $\gamma_N(\varepsilon)$ show that the transition is already encoded in a specific finite-size morphology of the entropy landscape. In particular, the curvature $\gamma_N(\varepsilon)$ develops a localized extremal structure whose position evolves systematically with system size and whose sharpening anticipates the nonanalytic behavior of the infinite system. Within MIPA, this structure is not interpreted as a generic finite-size fluctuation or as a crossover artifact, but as the finite-$N$ manifestation of the same collective reorganization that becomes singular in the thermodynamic limit. In this sense, the relevant object is not only the thermodynamic-limit singularity itself, but the entire sequence of entropy-derivative structures that leads to it.

This interpretation becomes concrete through the finite-size marker $\varepsilon_\star(N)$. Once extracted from the curvature morphology, this marker defines a critical trajectory that converges smoothly toward the exact critical energy. The important point is not merely that the extrapolation works, but that it starts from a structurally meaningful finite-$N$ object. The thermodynamic critical point is therefore recovered as the endpoint of an ordered family of finite-size reorganizations, rather than as a singular object that appears only at $N=\infty$. In this way, the finite system is not simply an imperfect approximation to the infinite one, but already contains the relevant critical organization in a measurable form.

A second important outcome of the present study is that this entropy-based organization is not isolated from the standard thermodynamic description. Once the entropy-derived trajectory is identified, independent observables organize coherently around it. The canonical caloric curve, the specific-heat anomaly, and the order-parameter sector all approach their asymptotic behavior in a way that is consistent with the same underlying trajectory. This coherence is essential. It shows that the finite-$N$ marker extracted from the entropy derivatives is not an artifact of a particular representation, but rather the microcanonical expression of the same collective rearrangement that other observables capture more indirectly. In this sense, the entropy derivatives do not compete with standard observables; rather, they provide the intrinsic structural level from which the latter may be reinterpreted.

From a broader perspective, the mean-field $\phi^4$ model plays the role of a stringent benchmark, not because it is the only system in which such structures arise, but because its thermodynamic-limit behavior is known analytically and can therefore be used as a precise reference. This makes the comparison especially transparent: the finite-$N$ entropy derivatives can be computed numerically and non-perturbatively, the exact infinite-size curves can be obtained independently, and the
corresponding critical trajectory can be tested quantitatively. In this sense, the present work is not simply another study of a mean-field transition. Its main purpose is to validate, in a controlled setting, the more general claim that criticality can be formulated as an intrinsic finite-size property through the morphology of microcanonical entropy derivatives.

The conceptual consequence is straightforward. Finite-size criticality should not be regarded as a blurred remnant of the thermodynamic limit. It is more natural to regard it as the regular finite-$N$ form of the same phenomenon whose singular version emerges asymptotically. In this language, extrema and inflection structures in entropy derivatives are not secondary numerical features; they are the operational signatures through which collective many-body reorganization becomes visible before any nonanalyticity can arise. The thermodynamic limit does not create criticality from nothing: it sharpens an already existing structural sequence until it becomes singular.

For the purposes of the present paper, this is the main conclusion. In the mean-field $\phi^4$ model, the microcanonical entropy derivatives provide direct and quantitatively controlled access to the finite-size organization of criticality. MIPA turns this organization into a unique finite-$N$ marker and a corresponding critical trajectory. The latter converges to the exact critical energy and simultaneously organizes the behavior of independent thermodynamic observables. Taken together, these results show that the transition is already present at finite
size as a structured property of the entropy landscape, while the thermodynamic singularity is its limiting expression. This establishes the mean-field $\phi^4$ model as a benchmark supporting a broader program in which phase transitions are identified not only through asymptotic singularities, but through intrinsic finite-size structures that can be reconstructed, classified, and followed across system size.

\begin{acknowledgments}
The calculations presented in this paper were carried out using the HPC facilities of the University of Luxembourg~\cite{VBCG_HPCS14} {\small (see \href{http://hpc.uni.lu}{hpc.uni.lu})} and those of the Luxembourg national supercomputer MeluXina.

The code used for the numerical simulations is publicly available at \cite{code_MF_phi4}.
\end{acknowledgments}


\end{document}